\documentclass[12pt,preprint]{aastex}

\newcommand{\kms}{km s$^{-1}\,$}
\shorttitle{EMILI}
\shortauthors{SHARPEE ET AL.}
\begin{document}
\slugcomment{To appear in \textsc{The Astrophysical Journal Supplement}}

\title{Introducing EMILI: Computer Aided Emission Line Identification}

\author{Brian Sharpee\altaffilmark{1}} \affil{Department of Physics
and Astronomy, Michigan State University, East Lansing, MI 48824}

\author{Robert Williams} \affil{Space Telescope Science Institute,
3700 San Martin Drive, Baltimore, MD 21218}

\author{Jack A. Baldwin} \affil{Department of Physics and Astronomy,
Michigan State University, East Lansing, MI 48824}

\and

\author{Peter A. M. van Hoof}
\affil{APS Division, Physics Department, Queen's University of Belfast,
Belfast BT7 1NN, Northern Ireland}

\altaffiltext{1}{Present Address: SRI International, 333 Ravenswood
Ave, Menlo Park, CA 94025}

\begin{abstract}
The identification of spectral lines can be a tedious process
requiring the interrogation of large spectroscopic databases, but it
does lend itself to software algorithms that can determine the
characteristics of candidate line identifications.  We present here
criteria used for the identification of lines and a logic developed
for a line identification software package called EMILI, which uses
the v2.04 Atomic Line List as the basic line database.  EMILI
considers all possible database transitions within the wavelength
uncertainties for observed optical emission lines and computes an
approximate intensity for each candidate line.  It searches for other
multiplet members that are expected to be seen with each candidate
line, and rank orders all of the tentative line identifications for
each observed line based on a set of criteria.  When applied to the
spectra of the Orion Nebula and the planetary nebula IC 418, EMILI’s
recommended line ID’s agree well with those of previous traditional
manual line assignments.  The existence of a semi-automated procedure
should give impetus to the study of very high signal-to-noise spectra,
enabling the identification of previously unidentified spectral lines
to be handled with ease and consistency.
\end{abstract}


\section{Line Identification In Rich Emission-Line Spectra}

Correct identification of spectral lines is fundamental to all
spectroscopic analyses.  For lines commonly observed in astronomical
spectra, a century of study has resulted in agreement on those
transitions that give rise to the stronger lines observed at visible
wavelengths.  However, there is still uncertainty about the proper
identification of many lines, and this problem is even more severe in
other wavelength regions.  As spectra achieve fainter detection
limits, the increasing number of transitions observed leads to a
larger fraction of uncertain identifications.  The effort involved in
making correct line identifications for the large numbers of lines
detected in high signal-to-noise (S/N) spectra can be daunting,
especially since identifications must be made on the basis of overall
astrophysical consistency, causing correct line identifications to be
mutually interdependent.  This problem has been studied in the past
for stellar absorption spectra, and techniques have been developed for
distinguishing between chance coincidences and true identifications
\citep{HCC73,CA90}.
   
Line identification is often problematic for emission-line objects
because, unlike absorption lines which are usually formed near
conditions of thermodynamic equilibrium, emission lines are formed
under conditions where coincidental resonances and unusual excitation
mechanisms cause isolated individual transitions to have strengths
that deviate from their equilibrium values by many orders of
magnitude.  The difficulty in making correct identifications for the
large numbers of faint lines observed in emission spectra has acted as
a disincentive to obtain the very high signal-to-noise spectra that
are necessary for their detection.  However, deep high resolution
spectra of both H II regions \citep{E98,E99,B00} and planetary nebulae
\citep{L95,L00} are now becoming routinely available. Valuable
information exists in the detection of previously unobserved faint
ionic species \citep{PB94}, so it is important to confront the
challenge of line identifications in an efficient and systematic way.
   
The usual approach to identifying emission lines in these high-quality
spectra has been to start with the line identifications available in
the literature for the spectra of similar objects. After that, it is
necessary to manually work through multiplet tables and other line
lists to try to arrive at identifications that make physical sense in
terms of wavelength agreement, line intensities, and the presence or
absence of other transitions within the same multiplet or from the
same ion. This procedure, which we will refer to as the “traditional”
approach, is both tedious and prone to being unsystematic.
  
One technique that is now being tried is to construct synthetic
spectra of the object under study and fit them to the observed
spectrum \citep{W01}, similar to well-developed methods for analyzing
absorption-line spectra. This technique has many strengths, including
a rigorous treatment of blended features and flexibility in dealing
with the wavelength uncertainties of the spectral database(s).  There
is no doubt that this procedure should figure prominently in efforts
to identify spectral lines.
   
In this paper we employ a different approach in which we describe a
semi-automated technique for identifying emission lines. The
centerpiece is a computer program called EMILI (EMIssion Line
Identifier). It automatically applies the same logic that is used in
the traditional manual identification of spectral lines, working from
a list of measured lines and a database of known transitions, and
trying to find identifications based on wavelength agreement and the
relative computed intensities of putative ID’s, on the presence of
other confirming lines from the same multiplet or ion.
   
We have developed EMILI in the context of analyzing high S/N echelle
spectra of planetary nebulae that can typically contain 500–-1000
emission lines, some down to an intensity level $10^5$ times fainter
than H$\beta$. Our interest is in measuring chemical abundances from
the faint lines of elements heavier than H and He. We have found that
these spectra include large numbers of emission lines for which atomic
parameters such as collision strengths and transition probabilities
are not accurately known, but we realize the importance of including
these lines in the analysis so that we can establish which ions are
represented in the spectrum.  EMILI is therefore designed in the
spirit of using rough, order-of-magnitude estimates of atomic
parameters. We believe that this is far better than ignoring such
lines when in fact they are seen in large numbers in the observed
spectra. Since EMILI’s goal is simply to give possible identifications
for lines, only a very crude model of the ionized nebula is
needed. EMILI works from an input list of the wavelengths and
intensities of the many hundreds of observed emission lines, and for
each line develops a short output list of suggested identifications,
rank-ordered in a preliminary way according to their plausibility. The
astronomer then reviews the output list and chooses the best
identification, based on physical insight.  Eventually, after we have
obtained sufficient experience with EMILI for different types of
objects and have incorporated additional criteria by which correct
line identifications can be discerned, the situation will evolve to
one in which line ID’s can be assigned automatically without the
necessity of insight.  EMILI is most beneficial in being applied to
spectra which reveal large numbers of lines not normally seen in
emission spectra.  Since the identifications of virtually all of the
stronger lines in most astronomical objects have long been known, this
occurs for (a) unusual types of objects, and (b) high S/N spectra for
which large numbers of faint lines are detected.

\section{Preliminary Data Reduction Steps}\label{steps}

As stated above, the input to EMILI is a long list of observed
wavelengths and intensities of emission lines. Before using EMILI,
these must be measured in some way from calibrated, co-added
spectra. Our technique and level of accuracy for reducing deep echelle
spectra are described by \citet{B00} and in a forthcoming companion to
the present paper \citep{SBW03}. Basically, we use a combination of
standard IRAF and specialized FORTRAN programs to extract
one-dimensional spectra, binned along the slit, from the
two-dimensional images that come from the spectrograph. The flux
calibration is determined by observing several standard stars through
a wide slit. The observational uncertainties are assessed from the
quality of the fits to the wavelength calibration and standard star
spectra, by comparing to previously-published results for the same
object, and by comparing overlapping parts of the observed spectrum
that come either from adjacent echelle orders or from different
grating setups. In our 9 \kms FWHM spectra, the wavelength accuracy is
typically 1 \kms and the line flux accuracy is 10–-20 percent for all
but the very weakest lines.
   
The next step is to detect and measure the emission lines contained in
these extracted spectra, down to as faint a level as possible. The
specification of what constitutes an emission line and what its
characteristics are is a critical part of line identification.  It is
very helpful if the object is spatially resolved and a 2-dimensional
spectrum is available to aid in distinguishing real lines from
artifacts, the latter of which contaminate the object spectrum and are
problematical at low signal-to-noise levels.  We use the rdgen
algorithm, which is part of the vpfit software package of
\citet{C01}\footnote{\url{http://www.ast.cam.ac.uk/$\sim$rfc/rdgen.html}}.
The rdgen program takes a calibrated spectrum and passes a window
along in wavelength to determine the flux and S/N at each wavelength.
The probability that a particular feature is an actual emission line
is then determined from criteria related to the local flux relative to
a fitted continuum level and the flux profile, e.g., line width.  The
flux, width, S/N, wavelength and its uncertainty are determined for
each line, and this information is then used to compile an observed
line list that serves as the basis for the identification
procedure. The major benefit of using rdgen is that it finds a
complete set of emission lines down to a consistent S/N limit, so that
we can assess the significance of the failure to find emission lines
that in principle should be present at some intensity level. However,
we caution that EMILI in its present form does not make use of the S/N
for measured lines or for the upper limits of unobserved lines.

\section{The EMILI Code}
 
EMILI is a stand-alone FORTRAN code that runs in 5–-10 minutes on any
UNIX, LINUX, or Windows computer that has a suitable FORTRAN
compiler. It is publicly available over the web
\citep{S03a}\footnote{\url{http://www.pa.msu.edu/people/sharpee/emili.html}},
with a primer, and has the following logical flow.  For each line in a
list of unidentified observed lines submitted to EMILI, a transition
database is queried for all transitions within the immediate
wavelength vicinity.  A separate list of pre-identified ``signature
lines'' from the same spectrum is used to establish kinematic and
ionization models of the observed object.  EMILI calculates a
predicted template flux for all candidate emission lines considered in
the line list based upon these models and upon the characteristics of
each transition.  For each candidate transition in the database EMILI
searches the line list to identify other transitions from the same
multiplet.  EMILI then ranks each candidate ID for an observed feature
according to wavelength agreement, strongest relative predicted flux,
and the numbers of multiplet members detected, and it presents the
results to the user for final ID determination.
   
In the following sections we specify in more detail the algorithms and
general approach used in EMILI.

\subsection{Input file of observed lines}

EMILI requires an ASCII input file containing the measured (a)
wavelength, (b) 1-sigma wavelength uncertainty due to measurement
error, (c) intensity relative to a fiducial line such as H$\beta$, (d)
line width (FWHM), and (e) signal-to-noise relative to the adjacent
continuum.  In our case the output file from rdgen, after repeat
measurements have been averaged together and night-sky lines and
obviously spurious lines (due to cosmic ray hits, etc.) removed,
serves as the input file for EMILI.

\subsection{Spectroscopic Database}
   
A second major input to EMILI is a database of atomic transitions that
are to be used as candidate ID’s for observed emission lines.  Recent
compilations of large electronic databases of transitions are what now
makes it practical to use computers to suggest line identifications
using the same logic that has been employed in the traditional
identification of lines.  Computer-aided identification is especially
valuable in facilitating a comparison of possible identifications for
a given line with the putative identifications for other lines.  The
key to the identification procedure is the database of transitions
used in the search process.  Fortunately, several extensive databases
have been developed in recent years which are accessible
electronically and which are continually being augmented as new data
are made available.  One of the most authoritative of these is the
NIST Spectroscopic
Database\footnote{\url{http://www.physics.nist.gov/cgi-bin/AtData/main\_asd}},
which consists largely of transitions that have been observed in
laboratory measurements.  The NIST transitions information is
generally quite reliable, although incomplete.  Some lines that are
observable in astrophysical spectra have been added, most notably
forbidden transitions, but still many transitions of ions do not
appear because confirming data are considered to be lacking by NIST
standards.  The incompleteness of databases, especially proper
wavelengths, is a problem for line identification for which there is
no alternative.  Incomplete information will always limit the
viability of making line identifications by any technique.  

Other line lists exist, and one of the most complete and inclusive of
these for UV/optical/IR wavelengths is the v2.04 Atomic Line List
compiled by
\citet{V99}\footnote{\url{http://www.pa.uky.edu/$\sim$peter/atomic/}}. This
list uses a very different approach in its construction.  It is based
on observed energy levels of ions rather than observed transitions.
This set of levels is supplemented with theoretical predictions and
Ritz extrapolations where it is meaningful to do so.  The actual line
list is constructed by a computer program that imposes a carefully
chosen set of selection rules to determine which levels have either
allowed, intercombination, or forbidden transitions connecting them.
The wavelengths of the lines, including an estimate for the
uncertainty, are calculated from the straight difference of the level
energies (Ritz wavelengths).  This procedure allows the line list to
be far more complete since the only requirement is that the upper and
lower level have been observed, which is less restrictive than the
requirement that the line itself has been observed.  This is
especially important in the infrared where very few laboratory
experiments have been undertaken.  One drawback of this approach is
that observed transitions without a proper spectroscopic
identification cannot be included in the line list.  However, in the
long run this situation will remedy itself once these lines are
identified and an updated term analysis becomes available.
   
Numerous spectroscopic databases exist, and it is preferable to
interrogate line lists that are as complete as reasonably possible
because a successful identification logic will reject specious
transitions.  As is done in traditional studies to identify spectral
lines, multiple sources that list valid atomic and molecular
transitions should be utilized when considering putative line
identifications in astronomical spectra.  However, for the initial
development of EMILI we have confined the present study to the use of
only one database, the v2.04 Atomic Line List, because of the
different formats used in the electronic databases, which would
require separate interrogation schemes.  We eventually intend to
extend the capabilities of the software to interrogate multiple line
lists.

\subsection{Signature Lines}
   
Many emission-line objects have a kinematical structure that
segregates lines from different ionization stages in velocity.
Additionally, the level of ionization can vary greatly from object to
object, affecting the relative intensities of lines from different
ions.  Since wavelength agreement and predicted intensity are
important criteria for making identifications, we define a set of
signature lines spanning a range of ionization stages whose ID’s are
reasonably secure, and we use these lines to establish radial velocity
corrections to determine the zero-velocity, or laboratory, wavelength
for each observed line and to find an approximate ionization
distribution for the object that is used to predict template fluxes
for candidate line ID’s.  This information is then used with generic
cross sections and spontaneous transition coefficients to compute a
rough template flux for every putative line ID that is considered from
the transition database.  The signature lines are identified manually
by traditional procedures at the beginning of the process.
   
\subsection{Identification Criteria}
   
Although there are clear criteria by which a possible line
identification can be rejected, there are no criteria by which a line
identification can be guaranteed to be correct.  Even a line such as
H$\beta$ has at times been ascribed to a feature that was later shown
to actually be due primarily to \ion{He}{2} 8-4.  So, astrophysical
consistency and reasonableness are important considerations when
assigning identifications, which mitigates against unexpected ID’s,
and final consensus is often achieved only after a body of data have
been gathered for a large group of similar objects.  For weak lines
from ions with few other lines present or detectable, some doubt may
persist about the correctness of an ID.
   
The criteria we have used in making emission-line identifications are:
(1) wavelength agreement, (2) the relative intensities of the
candidate transitions, as determined from an approximate calculation
using generic cross sections, and (3) the detection of other lines
from the same multiplet that are expected to be present with the
candidate line.  Based on the extent to which each candidate line
satisfies the above criteria a numerical value is assigned to that
transition, and a relative ranking of all reasonable ID’s from the
database is arrived at for each observed line.  Line ID’s are made on
the basis of this ranking.

\subsection{Velocity Shifts and Ionization}
   
Basic information about the spectrum that is necessary for normal line
identification procedures is obtained from the ''signature'' lines
that can be among the stronger of those observed in emission spectra.
These lines, which span a wide range in ionization, are searched for
and identified manually before the software is applied to the
spectrum.  The signature lines are used to determine the velocity
shift of the spectrum being studied, including differences in velocity
between lines of different levels of ionization, and an approximate
distribution in ionization of the emitting ions which is used to
calculate expected line intensities, i.e., template fluxes, of
candidate lines.  For most objects H and He are the dominant sources
of continuum opacity and therefore we specify levels of ionization
according to the ionization potentials of these elements.  We
arbitrarily establish five different levels of ionization, from very
low to very high, by defining the discrete bins that are specified in
Table~\ref{tbl1}.  Listed for each ionization bin are selected lines
from ions belonging to that bin, i.e., the signature lines.. The
observed intensities of the signature lines are proportional to the
fractional abundances of their parent ions, which pertain to the
ionization level of that bin, and their intensities are used to
determine the general ionization of the spectrum.  The fractional
abundances $x_k$ of ions in each energy bin are determined from the
intensities of the signature lines for each bin as follows.
   
Bin 1 represents those ions having ionization potentials less than
that of hydrogen, and although the intensities of lines such as
\ion{Mg}{1}], [\ion{S}{1}], [\ion{C}{1}], and \ion{Ca}{2} depend upon
the heavy element abundance and kinetic temperature, we determine $x_1$
independent of these parameters in the following manner.  Let $F_1$ be
the flux of the strongest of the signature emission lines for Bin 1.
Then,
\begin{equation}
\label{eq1}
x_1 = \left\{ \begin{array}{c} 10^{-3} \\ 10^{-2} \\ 10^{-1} \end{array}
\,\, \textrm{when} \,\, F_1/F_{\textrm{H}\beta} = \left\{\begin{array}{c}
< 10^{-4} \\ 10^{-4}-10^{-2} \\ > 10^{-2}
\end{array}\right. 
\right. \,.
\end{equation}

The ionization correction factors for moderately ionized species are
determined from the relative strengths of the \ion{He}{1} lines compared to
H$\beta$, which depend on an assumed helium abundance (default is solar)
through the relation from recombination theory that
\begin{equation}
\label{eq2}
\frac{x_3}{x_2} = 0.7\, Y \frac{F_{\lambda5876}}{F_{\textrm{H}\beta}}
= 2.0\, Y \frac{F_{\lambda4471}}{F_{\textrm{H}\beta}} \,,
\end{equation}
where $Y$ is the He/H abundance by number. 
   
The fractional ionization of more highly ionized ions is obtained from
the intensity of \ion{He}{2} $\lambda$4686 relative to the \ion{He}{1}
lines through the relations
\begin{equation}
\label{eq3}
\frac{x_4}{x_3} = 0.04 \frac{F_{\lambda4686}}{F_{\lambda4471}} = 0.11
\, \frac{F_{\lambda4686}}{F_{\lambda5876}}\,.
\end{equation}
    
Finally, the ionization correction factors for the very highest
ionization levels (I.P. $>$ 100 eV) are determined from the
intensities of lines such as [\ion{Ne}{5}], [\ion{Fe}{7}],
[\ion{Fe}{10}], and [\ion{Ar}{10}] via the relation
\begin{equation}
\label{eq4}
x_5 = 10^{-3} + \frac{F_5}{F_{\textrm{H}\beta}}\,,
\end{equation}
up to a maximum value of $x_5 = 0.3$, where $F_5$ is the flux of the
brightest of the signature lines for Bin 5 (see Table~\ref{tbl1}).
The above relations for the $x_k$, together with the condition that
$\sum x_k = 1$ when summed over all of the ionization bins, specify
the ionization level of the spectrum.  In cases where no signature
lines are observed for a particular ionization bin, EMILI sets a
minimum value of $x_k = 10^{-3}$.  Once the general ionization
distribution for the spectrum is determined from the above relations,
the relative ion abundance for specific elements is arrived at in the
following manner.  Designate the lower and higher ionization energy
limits for each bin $k=1-5$ by $E_{k-1}$ and $E_k$, and the fractional
abundance of ions associated with that bin as $x_k$.  Designate the
ionization potential of ion i and that of its next lower stage of
ionization as $\chi_i$ and $\chi_{i-1}$.  When an ion $i$ and its next
lower stage of ionization fall within the same energy bin, i.e., when
$E_{k-1} \le \chi_{i-1} < \chi_i \le E_k$, set $x_i = x_k$.  However,
when two consecutive ionization stages fall into different energy
bins, e.g., when $E_{k-1} \le \chi_{i-1} \le E_k$, and $E_k \le \chi_i
\le E_{k+1}$, EMILI sets $x_i = (x_k + x_{k+1})/2$.  For the special
cases of H and He, $x(\textrm{H}^+)=x_2$, $x(\textrm{He}^+)=x_3$, and
$x(\textrm{He}^{+2})=x_4$.  Although ionization fractions determined
this way are only approximate they are adequate for order-of-magnitude
intensity calculations for lines from different ions.

\subsection{Template Fluxes}

One of the obvious criteria for making line ID’s, especially useful
for distinguishing between transitions that have essentially the same
wavelength, is the expected flux of each candidate line ID compared to
the intensity of the observed line.  The excitation mechanisms and
relevant cross sections for each transition are required to compute
its expected intensity, and these are not known for the vast majority
of lines.  However, for purposes of dealing with large numbers of
lines, generic cross sections can be used and the excitation processes
that are common for most observed lines can be assumed to operate for
all levels.  These assumptions can be substantially in error for
individual transitions, but for purposes of helping to distinguish
between the relative strengths of transitions of different ions such
calculations should have some validity in a statistical sense when
applied to large numbers of transitions.  

For nebular conditions, i.e.\ low density gas in a dilute radiation
field, excitation is normally caused by electron impact from the
ground state and electron recapture from the next higher stage of
ionization.  Near a strong continuum source absorption by resonance
transitions followed by cascading can also produce line emission.
Cross sections for each of these processes have been calculated for
numerous levels of many ions, and they have dispersions of several
orders of magnitude for different levels.  Thus, it is easy for the
predicted, or template, flux of a transition to be in error by factors
of 100 when generic cross sections are used.  Nevertheless, if the
relative abundances of the different ions are known, the predicted
fluxes of two competing candidate transitions of widely different
abundance or excitation level still can be a telling criterion for
preferring one line over the other as a putative identification for an
observed feature.

We use a simple approximation to compute the template flux, $F_t$, of
emission lines associated with each and every transition in the
database.  We consider all emission lines to be excited by both
collisional excitation and recombination processes, representing their
contributions to the flux of any line from ion $i$ by the expression
\citep{O89},
\begin{equation}
\label{eq5}
F_t= A \left( x_i \frac{\exp(-0.8 \chi_j)}{1+K_j\,,n_e} + 10^{-5}
x_{i+1}(i+1)^{1.7} C_j\right)\,,
\end{equation}
where $A$ is the element abundance relative to H, $x_i$ and $x_{i+1}$
are the fractional abundances of the ions $i$ and $i+1$, $\chi_j$ is
the excitation potential of the upper level of the transition in eV,
and $n_e$ is the electron density of the gas (cm$^-3$).  The term with
constant $K_j$ accounts for collisional de-excitation of low-lying
levels, and the constant $C_j$ is proportional to the transition
probability of the line.  Both constants take on values that depend on
the type of transition, such that for (a) permitted electric dipole
transitions, $K_j =10^{-14}$ and $C_j =1$; for (b) electric dipole
intercombination, or spin forbidden, transitions, $K_j =10^{-9}$ and
$C_j =10^{-4}$; and (c) all other types of transitions, e.g., magnetic
dipole and electric quadrupole, $K_j =10^{-6}$ and $C_j = 10^{-7}$.
Eqn~\ref{eq5} predicts an approximate relative flux for any transition
under typical nebular conditions.  All line intensities so calculated
are normalized to the H$\beta$ flux predicted from the same
expression, and are referred to as the template fluxes of the database
lines.

\subsection{Associated Multiplet Lines}
   
The presence of other lines originating from the same upper level or
from within the same multiplet is one of the more useful criteria by
which line identifications can be judged.  Although multiplets are
defined by the coupling scheme appropriate for the ion, except for
very level-dependent excitation processes involving resonances one
generally expects for a given ion that lines originating from levels
of similar excitation potential tend to be present with similar
intensities.  This is especially true within individual multiplets.
Most of the more abundant elements have low atomic number and the
stronger optical transitions of many of the ions of these elements
tend to obey LS- or jK-coupling, so the multiplets that are most
likely to be present in astronomical spectra can generally be clearly
specified.  If experience shows that this assumption is too frequently
violated, different methods for determining associated transitions may
be considered.
   
The current EMILI algorithm will determine for all possible
LS-coupling transitions in the database other members of the same
multiplet that are expected to be present with intensities similar to
that of the primary transition.  Since relevant atomic data are not
known for the vast majority of transitions, we rely upon general
principles.  Additionally, all multiplet lines grouped within the
instrumental resolution or natural line width are considered to be a
single line.
   
Level populations and spontaneous transition rates within a multiplet
tend to be larger for those lines originating from upper levels with
the highest statistical weights.  We determine for every transition
those lines within the same multiplet that are expected to be
observable at intensities comparable to or greater than its flux.  We
call these lines within the multiplet the ``associated lines'' of the
candidate line (or putative ID), and we arbitrarily define them to be
those lines within the multiplet originating from upper levels with
$J^{\prime} \ge J_{u-1}$ and ending on lower levels $J^{\prime\prime}
\ge J_{l-1}$, where $J_u$ and $J_l$ are the angular momenta of the
upper and lower levels of the line under consideration.  This
definition may be unnecessarily restrictive, especially in its
limitation on the lower levels of the associated transitions, but we
wish to err on the side of considering those multiplet members that
are most likely to have intensities comparable to the candidate line.
The detectability of associated lines is also dependent upon the
signal-to-noise of the lines and is affected by chance coincidences
and line blends, so the presence or absence of associated lines as a
constraint for identification of a line has limitations, but the
general concept is an important one to invoke for the validation of
line identifications.

\subsection{Numerical Identification Index}\label{multi}
   
We base all line identifications on the three criteria discussed
above: wavelength agreement, strongest computed template flux, and
presence/absence of associated lines from the same multiplet.  In
order to put line identification on a quantitative basis we establish
a numerical identification index (IDI) that assesses the extent to
which every putative line ID for an observed line satisfies the
criteria.  Since the three criteria are independent of each other,
separate numerical values are defined for each of the individual
component criteria, and the IDI is defined as the sum of the three
components.  For the present we arbitrarily assign numerical values to
how well candidate lines satisfy each of the criteria, however in the
future it might be instructive to weight each component in such a way
that the line ID’s suggested by the resulting IDI produce the best
agreement with previous published work.  Of course, there is no
guarantee that identifications in previous studies are correct.

The ID Index which we have instituted for EMILI is defined to be
\begin{equation}
\label{eq6}
\textrm{IDI} = W + F + M \,, 
\end{equation}
where $W$, $F$, and $M$ are the wavelength, flux, and multiplet components,
respectively, of the IDI, and each take on integer values between 0
and 3, with lower scores being better, according to the following
conditions.

\noindent
a) Wavelength Component
   
Define $\lambda_o$ to be the wavelength of an observed line corrected
for the object radial velocity, and $\lambda_l$ to be the wavelength
of a candidate line from the database corrected for any
ionization-dependent velocity shifts deduced from the signature lines.
Let $1\sigma$ be the standard deviation in the measured wavelength of
the observed line.  Then, $W$ = 0, 1, 2, or 3 for candidate lines for
which $\left|\lambda_o - \lambda_l\right| \le 0.5\sigma\,,
1.0\sigma\,, 1.5\sigma$, and $2.0\sigma$ respectively.  Uncertainty in the
laboratory wavelengths are not taken into account in this
determination, although consideration will be given to doing so in the
future.

\noindent
b) Flux Component 

Designate the computed template flux of a candidate line by $f_1$, and
let $f_b$ designate the brightest template flux of all candidate lines
within 5$\sigma$ in wavelength of the observed line.  Then, $F = 0$
for that line having the brightest predicted flux if $f_b > 10 f_1$
for all other candidate lines within that wavelength interval.
Otherwise, $F$ = 1, 2, or 3 for lines having $f_1 \ge 0.1 f_b\,, 0.01
f_b\,,$ and $0.001 f_b$, respectively.  For lines fainter than
$10^{-3} f_b\,, F = 4$.  Thus, the flux component of the ID Index
takes into consideration comparison of the predicted template fluxes
of the candidate lines with each other, but not with the observed flux
of the line.

c) Multiplet Component 

For each candidate line from the line list designate $P$ as the number
of associated multiplet lines, as defined above in \S~\ref{multi}, for
that line.  Define $D$ to be the number of associated multiplet lines
that appear to be detected, i.e., for which a line is observed at the
appropriate wavelength and having a flux within an order of magnitude
of the primary candidate line.  Then, (i) $M = 0$ when $P:D$ = 1:1, or
when $D \ge 2$.  (ii) $M = 1$ when $P:D$ = 0:0 or 2:1.  (iii) $M = 2$
when $P:D$ = 1:0 or ($\ge$3):1.  And, (iv) $M = 3$ when $P:D$ =
($\ge$2):0.  For every observed feature in the spectrum the master
line list is searched for possible ID’s within a specified wavelength
range, typically $\pm±3\sigma$ of the wavelength of the observed
feature, and the ID Index is determined for each candidate line.
Identifications are assigned on the basis of the IDI, with lower
values of IDI signifying a higher probability of correct
identification.

\section{An Example: Application Of EMILI To IC 418}

We have undertaken a program to obtain high dispersion, high
signal-to-noise spectra of a few selected PNe because they are among
the best objects to observe for the detection of faint emission lines.
The primary motive has been to identify as many CNONe recombination
lines as possible in order to compare the relative intensities of
these lines between themselves and with the strong forbidden lines
from object to object.  Data obtained and reduced for the relatively
low ionization PN IC 418 using the CTIO Blanco 4m telescope + echelle
at a spectral resolution of 33,000 over the wavelength range of
3400-9700 \AA\ are described in \citet{SBW03}.  We have taken the list
of emission lines defined by rdgen as applied to the spectrum in that
paper and have applied EMILI to the line list using the procedure that
has been outlined in \S~\ref{steps} above, and which is also described
on the EMILI website.  For a line to be considered real we require
S/N$>$7, with the exception of 23 features in the range $7>$S/N$>3$
which were deemed real lines upon inspection of the original
2-dimensional spectra images.
   
Line identifications are made on the basis of the ID Index defined in
eqn~\ref{eq6}, with the most probable ID taken to be that line among the
candidates considered that has the smallest value of the Index.  In
order to present the information used to compute the Index for every
candidate ID, for each observed feature EMILI lists all reasonable
ID’s for that line together with the wavelength, predicted flux, and
associated multiplet lines for each candidate ID.  The output table
for EMILI thus consists of a list of every emission line that is
observed in the spectrum, as defined by rdgen, together with the
possible transitions (and their characteristics) that might be
identified with that observed feature.
   
\begin{figure}
\plotone{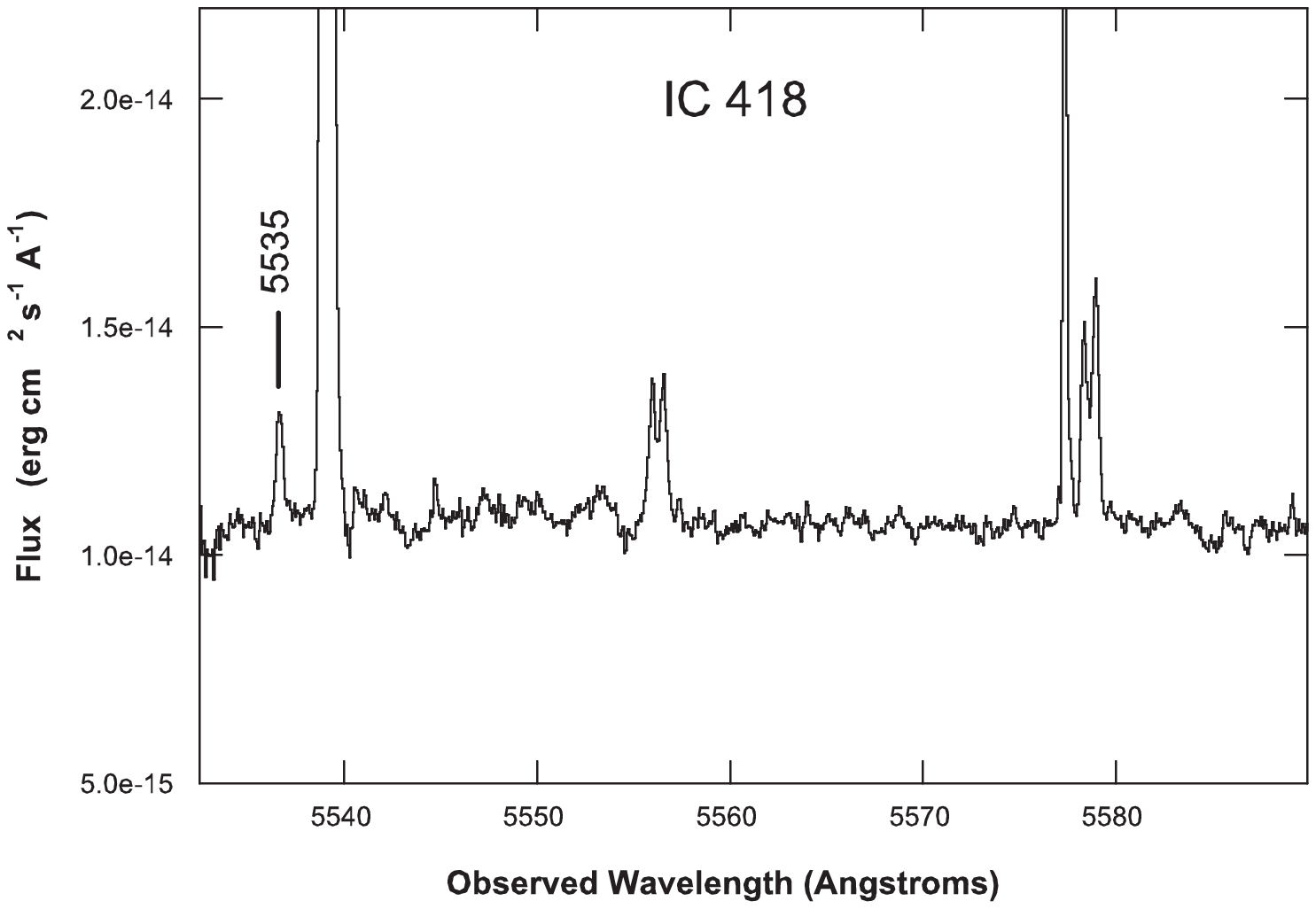}
\caption{A portion of our echelle spectrum of IC 418, showing the
emission line at 5535\AA\ that is used as an example in
Table~\ref{tbl2}. \label{fig1}}.
\end{figure}

As an illustration of EMILI output and results, we consider the
identification of an emission line observed at 5536.60\AA\ in the
spectrum of IC 418, with the relevant EMILI output for this line
listed in Table~\ref{tbl2}.  The data were obtained with the CTIO 4m
echelle in Dec 2001 at a resolution of 9 \kms (R=33,000), and the
relevant wavelength region of that spectrum is shown in
Fig.~\ref{fig1}.  The measured wavelength of the line is 5536.60\AA,
which when corrected for the +68.6 \kms radial velocity of the object
determined from the higher Balmer and Paschen lines, corresponds to a
rest wavelength of 5535.33\AA.  Its observed flux relative to H$\beta$
is $4.5 \times 10^{-5}$, and the line has a signal-to-noise ratio of
S/N = 26 and a width (FWHM) of 17 \kms.  These measured line
attributes appear in the first row of Table~\ref{tbl2}, and below this
row in Columns A-K appear all lines listed in the v2.04 Atomic Line
List that have wavelengths within 5$\sigma$ (about 20 \kms or 0.37\AA)
of the observed line and which have template fluxes within a factor of
$10^3$ of the brightest computed template flux for the entire group of
candidate lines.
   
Column A lists the observed wavelength (air) of the unidentified lines
corrected for any velocity shifts appropriate for the emitting ion of
each candidate ID, according to the kinematical model. Transitions
whose wavelengths are denoted by a ``+'' sign are within $1.5\sigma$
wavelength error from the measured line wavelength.  The laboratory
wavelength (in air) is given in Column B, and the emitting ion is
listed in column C.  Columns D and E give the predicted template flux
for each candidate line and the difference between its wavelength and
that of the measured line in units of velocity (\kms).  Column F lists
the number of associated multiplet lines that should be observable
compared to the number observed.  In column G the Identification Index
is given for each candidate line, and the capital alphabet letter
following the numerical IDI gives the ranking of the line, with A
representing the most likely ID, i.e., the lowest IDI value.  Finally,
in columns H/I/J/K appear the wavelengths of the strongest associated
multiplet lines that are possibly observed together with their
differences in wavelength (in \kms) from those of the observed lines.
Our experience shows that when the associated lines are truly that,
and not just coincidences, their differences in wavelength from the
observed lines are virtually identical to the difference between the
primary line and its measured line wavelength.
   
Looking in detail at the EMILI results for the observed IC 418
λ5536.60\AA\ line, a secure identification with \ion{N}{2}
$\lambda$5535.35\AA\ is indicated, although an ID with \ion{C}{2}
$\lambda$5535.35\AA\ is also a possibility.  The \ion{N}{2} line has a
slightly (insignificantly) better wavelength agreement with the
observed line than the \ion{C}{2} transition, although the \ion{C}{2}
line is predicted to be slightly brighter than the \ion{N}{2} line
using the generic cross sections and abundances.  Both putative ID’s
have computed template fluxes that are higher than that of the
observed line.  The key to the identification devolves to the
associated multiplet lines for the two transitions: both possible
lines in the \ion{N}{2} multiplet are apparently present whereas the
one possible associated line for the \ion{C}{2} line is not observed.
The former lines could conceivably be due to chance coincidences with
unrelated transitions, however the wavelength differences between the
\ion{N}{2} associated line wavelengths and those of the observed lines
are very similar: -1.0 \kms for the primary line vs.\ –2.9 and –1.3
\kms for its associated multiplet lines, which argues against chance
coincidences.  The lack of detection of the \ion{C}{2} associated
multiplet line could be due to a number of factors having nothing to
do with its true intensity, including its location at the very edge of
an echelle order or its superposition on a strong night sky line or
scattered ghost feature.  Consequently, its non-detection may be
explainable.  These doubts can be addressed by visual inspection of
both the original 2-dimensional spectral image and the final reduced
1-dimensional spectrum.  This final manual check of the EMILI results
is an important component of proper line identification when there are
several competing transitions that are credible ID’s.
   
It is worth noting that the particular \ion{N}{2} $\lambda$5535.35\AA\
transition discussed above is a quartet line whose upper level is an
autoionizing state that lies above the N$^+$ ionizing continuum.  It
is therefore almost certainly excited by dielectronic recombination of
N$^{+2}$.  Of particular significance is the fact that all three
possible stabilizing transitions from the autoionizing state are
observed in IC 418, making their identification quite secure.
   
We have applied EMILI to the full, final reduced high S/N echelle
spectrum of IC 418 obtained at CTIO.  We employed an updated version
of the v2.04 Atomic Line List that includes higher level lines of
\ion{He}{1}, and a standard set of parameters in computing template
fluxes for putative line ID’s, i.e., solar abundances, and $n_e=10^4$
cm$^{-3}$ and $T_e=10^4$ K.  In making final ID’s for this nebula we
have used the EMILI results as the initial basis for considering final
line assignments, however we have not blindly accepted the EMILI
recommendation for each line.  Rather, we have studied the entire
spectrum and have considered the entire list of ID’s collectively,
using our judgment as to what lines we believe constitute the most
reasonable identifications, and these are presented for the entire
spectrum in Table~\ref{tbl3}.
   
In most cases we have accepted the EMILI top-ranked ID as the final
ID.  However, for some lines we have selected one of the ID’s whose
IDI was not the smallest of the candidate group, as evidenced by the
ranking given in column (6) of Table~\ref{tbl3}.  For almost every
line our final ID was one that was ranked by EMILI as one of the four
most probable lines, and we have listed all the final identifications
together with the observed lines and their measured wavelengths and
reddening corrected fluxes in the Table.  When the lowest value of IDI
for an assigned line is higher than IDI$\ge$5 we consider the ID to be
uncertain and therefore tag that ID with a colon.  When the most
likely putative ID has IDI$\ge$8 we place a ``?'' after the ID,
believing that the ID does not have a solid basis and that the
spectral feature may be spurious or the line list does not contain the
correct transition for that feature.  This line list should constitute
one of the most detailed emission spectra of any PNe, and can serve as
an archetype of low ionization spectra, similar to the spectrum of the
Orion Nebula presented by \citet{B00}.
   
EMILI found solid identifications for 620 of the 807 observed IC 418
emission lines, and possible identifications for an additional 72
lines. Table~\ref{tbl3} notes for each line those identifications that
are not rated as questionable. There are a total of 750 such line
identifications, for 476 observed lines for which there was only one
suggested identification and for a further 144 observed lines for
which there is more than one possible identification. The dereddened
strengths of these identified lines range down to slightly below
$10^{-5}$ the intensity of H$\beta$.

\section{Comparision Of EMILI Results With Previous Studies}

The correctness of spectral line identifications, especially for
fainter lines that are not frequently observed, is difficult to
ascertain.  There is no absolute benchmark of correct ID’s with which
to compare the results of EMILI, except possibly for the stronger
lines in spectra which have been observed in many objects and for
which there is universal agreement.  We will therefore undertake to
compare EMILI results with those of previous studies done
traditionally as an indication of their reliability.
   
We are currently engaged in a program to obtain high S/N spectra of a
sample of emission-line objects to which we can apply EMILI.  At
present we have excellent data for IC 418 \citep{SBW03}, and we have
access to high resolution spectra of the Orion Nebula \citep{B00}, for
which detailed traditional line identifications have also been made.
For IC 418 we compare the EMILI identifications from this study with
those made by \citet{HAF} from their Lick Observatory echelle data.
   
We have taken the spectra of IC 418 and the Orion Nebula, and have
generated the necessary line characterization and wavelength error
tables from rdgen.  This information has been fed into EMILI using our
updated v2.04 version of the line database and standard nebular
parameters with solar abundances.  Line identifications have been
made, and when the lowest value of the IDI has been shared by more
than one line all of those lines are assigned as ID’s.  The resulting
identifications have then been compared with the published ID’s for
the Orion Nebula and IC 418.  The comparison of EMILI ID’s with those
assigned traditionally for these nebulae is given in Table 4.  The
EMILI ID rankings A, B, C, \& D refer to the first, second, third, and
fourth highest ranked ID’s from the algorithm.  The comparison shows
that the EMILI identifications ranked as ``A'' agree with the
traditional manual ID for about 85\% of all lines.  Furthermore, 90--98
\% of all manual ID’s are ranked as A, B, C, or D by EMILI. The
agreement with those ranked A can be improved upon easily by
optimizing the way in the ID Index is defined.  We have taken a close
look at some of the disagreements between the manual ID’s and those
ranked ‘A’ by EMILI, and we believe that EMILI is more likely to be
correct than the traditional line identifications in a majority of
cases.  This raises the question as to which identification process,
traditional or software, yields more correct results.  This question
can only be answered after a larger sample of objects have been
studied at fainter flux levels so that some consensus emerges as to
the correct identification of the weakest lines.

\section{Summary and Future Prospects}
   
The time- and labor-intensive method of traditional identification of
spectral lines has acted as a deterrent to obtaining very high S/N,
high resolution spectra of astronomical objects.  The task of making
manual identifications for large numbers of lines has required such
effort that the focus has been on utilizing only well-known stronger
lines as diagnostics to determine physical conditions.  Yet, modern
spectrographs and detectors make very faint lines observable with only
a modest investment of telescope time, and the large amount of new
information that is certain to be contained in previously unstudied
weak lines can now be tapped.
   
Even with its current simple logic EMILI works well in identifying
lines in nebular spectra, and certainly well enough to justify
continued improvement.  It would benefit from new features such as the
ability to interrogate multiple line databases, and using additional
criteria to evaluate candidate ID’s, including a more sophisticated
way to deconvolve line blends.  Also, it should be applied to the
spectra of additional types of emission-line objects.  One of the
eventual goals is to include very heavy elements in the line lists so
that s- and r-process elements can be identified when lines from these
elements are detected in spectra.  This may require detecting lines
down to $10^{-6}$ the intensity of H$\beta$.  This flux level is 100 times
fainter than the nebular continuum of the emitting gas upon which the
lines are superposed, however the requisite S/N is achievable.  The
very large numbers of emission lines that would be revealed in such
spectra are precisely the situation that requires an automated line
identification aid such as EMILI.
   
The logic incorporated into EMILI is based upon traditional procedures
used by spectroscopists for making manual line identifications in
astronomical spectra.  There are a number of uncertainties involved in
the application of EMILI to spectra, and a number of areas in which
improvements should eventually be made.  One of the major difficulties
of spectral line identification is accurately characterizing blends of
multiple lines.  It is often difficult to prove whether a feature is
or is not a blend, and what the component wavelengths and line widths
are when it is a blend.  Emission-line objects can have a complex
kinematical structure that results in complicated line profiles which
differ from ion to ion of the same element.  For example, a line of
sight that passes through an expanding shell, as is the case for the
object IC 418, produces a distinct double-peaked profile that is
easily misinterpreted as two separate lines rather than a single
feature.  The stronger signature lines in the spectrum easily remove
this uncertainty because they generally do not suffer significant
blending and can therefore be used to characterize the different line
profiles.  For the current version of EMILI only a cursory effort was
made to use differing intrinsic line widths and profiles as
identification discriminants because of the complexity involved in
doing proper line deconvolutions for large numbers of lines.  However,
these criteria should be developed further in future versions of
EMILI.
   
One of the most important aspects of faint line detection is the use
of spectral resolution that can resolve the narrowest lines in the
spectrum, which is set by the thermal widths of the lines (typically,
of order 10 \kms) and bulk motions in the gas.  Higher spectral
resolution increases the accuracy of measured line wavelengths, which
is the most important criterion in making line ID’s, and it serves to
increase the contrast between peak line intensity and the surrounding
continuum.  The detectability of emission lines is further enhanced by
low bulk velocities of the gas and by the faintest possible continuum
against which the lines are detected.  One cannot escape the
bound-free and 2-photon continuum emitted by the ionized gas, but one
can select objects of low dust content in order to minimize the
contribution of the scattered continuum from the ionizing star.  The
number density of lines increases at fainter intensity levels and the
proper identification of the weakest lines becomes increasingly
uncertain due to the less well determined wavelengths of these
features and the higher fraction of associated multiplet lines that
are not observed because their S/N is below the threshold of
detection.  This is inevitable, and there will always be a substantial
fraction of lines at the threshold of detection for which the criteria
for confirmation are not well determined.  Line ID’s for the faintest
lines are generally less certain, and further progress can be made by
pressing for spectra with yet higher S/N, since the faintest flux
levels we have achieved do not approach a line density that is so high
that weak lines blend into each other to form a pseudo-continuum.
   
Finally, the v2.04 Atomic Line List is the only database that we have
interrogated in searching for potential identifications in the present
version of EMILI, and its wavelengths are accepted as valid for each
transition.  A more thorough search process should eventually
interrogate multiple line lists and should include molecular lines,
especially those that are known to be present in the night sky
spectrum \citep{O96}.  Imperfect sky subtraction inevitably results in
night sky lines being present in the spectra of many objects, and even
though these lines are not intrinsic to the object they do require
proper identification.
   
The current line identification software is still in a rudimentary
form, yet it is already shown to be reliable and efficient in making
an arduous process consistent and tractable. EMILI produces the
largest gains over the traditional method in situations where there
are large numbers of lines to be identified, as in our deep echelle
spectra of IC418. For spectra of lower resolution the current version
of EMILI is likely to be less successful because of the importance
that it assigns to wavelength accuracy.  We are experimenting with
making the relative weights given to the identification criteria
variable and dependent upon the resolution and S/N of the spectrum.
Further refinement of the software logic and steady progress in
compiling more complete transition databases should produce a reliable
product, and have a dramatic impact on high S/N spectroscopy in the
next decade.

\acknowledgements

 PvH thanks the Engineering and Physical Sciences Research Council of
 the United Kingdom for financial support.


\clearpage

\clearpage

\begin{table}
\caption{Signature Lines for Ionization Bins. \label{tbl1}}
\begin{center}
\begin{tabular}{ccl}\tableline\tableline
\multicolumn{1}{c}{Bin} & \multicolumn{1}{l}{Energy Range (eV)} & \multicolumn{1}{c}{Signature Lines} \\ \tableline
A & 0-13.6 & Mg I] $\lambda 4571$, Na I $\lambda 5890,5896,$ \\
& & [S I] $\lambda 7775$, [C I] $\lambda 8727$ \\
& & Ca II (H\&K) $\lambda 3934,3968$ \\
B & 13.6-24.7 & H$\beta$ \\
C & 24.7-54.5 & He I $\lambda 5876,4471$ \\
D & 54.5-100.0 & He II $\lambda 4686$ \\
E & $>100$ & [Fe X] $\lambda 6375$, [Ne V] $\lambda 3426$, \\
 & & [Fe VII] $\lambda 6087$, [Ar X] $\lambda 5533$ \\ \tableline
\end{tabular}
\end{center}
\end{table}

\clearpage

\begin{table}
\caption{Sample EMILI Output. \label{tbl2}}
\begin{center}
\begin{tabular}{rlllcrclcrcr}\tableline\tableline
\multicolumn{12}{l}{Observed Line: 5536.60  Flux: 4.5E-05 S/N: 26.40
FWHM: 16.9} \\ 
\multicolumn{1}{c}{\textbf{A}} &
\multicolumn{1}{c}{\textbf{B}} & 
\multicolumn{2}{c}{\textbf{C}} &
\textbf{D} & 
\multicolumn{1}{c}{\textbf{E}} &
\multicolumn{1}{c}{\textbf{F}} &
\textbf{G} &
\multicolumn{1}{c}{\textbf{H}} & 
\multicolumn{1}{c}{\textbf{I}} &
\textbf{J} &
\multicolumn{1}{c}{\textbf{K}} \\ \tableline

  5535.36 &  5535.040  & P  &  IV &  1.3E-06 & 17.4 & 2/0 & *  &          &       &          &      \\
  5535.33 &  5535.169  & Mn &  II &  9.4E-06 &  8.6 & 5/1 & 9  & 5764.280 &   8.7 &          &      \\
+ 5535.36 &  5535.325  & Ni & III &  5.4E-06 &  1.9 & 5/0 & 7D &          &       &          &      \\
+ 5535.33 &  5535.347  & N  &  II &  3.1E-04 & -1.0 & 2/2 & 1A & 5530.242 &  -2.9 & 5551.922 & -1.3 \\
+ 5535.33 &  5535.353  & C  &  II &  1.1E-03 & -1.4 & 1/0 & 3B &          &       &          &      \\ 
+ 5535.33 &  5535.383  & Fe &  II]&  1.6E-06 & -3.0 & 7/0 & 7D &          &       &          &      \\
+ 5535.33 &  5535.384  & N  &  II &  3.1E-04 & -3.1 & 8/1 & 4C & 5526.234 &  -3.1 &          &      \\  
  5535.28 &  5535.378  &[V  &  II]&  8.2E-05 & -5.1 & 7/0 & 7D &          &       &          &      \\
  5535.36 &  5535.476  & Fe & III &  9.9E-05 & -6.2 & 8/0 & 8  &          &       &          &      \\
  5535.28 &  5535.413  & Fe &  I] &  2.5E-06 & -7.0 & 5/0 & 9  &          &       &          &      \\
  5535.28 &  5535.418  & Fe &  I  &  5.6E-05 & -7.3 & 5/0 & 8  &          &       &          &      \\
  5535.33 &  5535.466\$& Ar &  II &  1.0E-05 & -7.5 & 0/0 & 7D &          &       &          &      \\
  5535.28 &  5535.455  & Mg &  I  &  3.5E-05 & -9.3 & 0/0 & 7D &          &       &          &      \\
  5535.33 &  5535.525\$& S  &  II &  4.9E-05 &-10.7 & 0/0 & 7D &          &       &          &  \\ \tableline     
\end{tabular}
\end{center}
\end{table}

\clearpage

\begin{deluxetable}{p{0.47in}rrrrp{0.19in}p{0.3775in}p{0.5325in}rp{1.415in}p{1.415in}p{0.2525in}p{0.2525in}p{0.2825in}}
\tablewidth{8.765in}
\setlength{\tabcolsep}{0.035in}
\tabletypesize{\scriptsize}
\rotate
\tablecolumns{14}
\tablecaption{EMILI Line Indentifications for IC 418 \label{tbl3}} 
\tablehead{
\colhead{$\lambda_o$} &
\colhead{FWHM} &
\colhead{$F(\lambda)/F(\textrm{H}\beta)$} & 
\colhead{$I(\lambda)/I(\textrm{H}\beta)$} & 
\colhead{S/N} & 
\colhead{IDI} & 
\colhead{ID} & 
\colhead{$\lambda$} & 
\colhead{$\delta$V} & 
\colhead{Lower Term/Config} & 
\colhead{Upper Term/Config} & 
\colhead{$J_l$} & 
\colhead{$J_u$} & 
\colhead{Multi} \\
\colhead{(\AA)} & 
\colhead{(km/s)} & 
\colhead{($F(\textrm{H}\beta)=100$)} &
\colhead{($I(\textrm{H}\beta)=100$)} &  
 & & & \colhead{(\AA)} &
\colhead{(km/s)} & & & & & \\
\colhead{(1)} & \colhead{(2)} & \colhead{(3)} & \colhead{(4)} & \colhead{(5)} & \colhead{(6)} & \colhead{(7)} & \colhead{(8)} & \colhead{(9)} & \colhead{(10)} & \colhead{(11)} & \colhead{(12)} & \colhead{(13)} & \colhead{(14)}
}
\startdata 
 3512.511 &  22.9 &   0.1026 &   0.1365 &     37.6 & 2  A &   He I    & 3512.505   &  -0.5 & 3Po 1s.2p                           & 3D 1s.12d                           &  2.0 & **** &      0/0\\ 
 3530.499 &  24.6 &   0.1509 &   0.2002 &  99990.0 & 3  A &   He I    & 3530.482   &  -1.5 & 3Po 1s.2p                           & 3D 1s.11d                           &  2.0 & **** &      1/0\\ 
 3554.417 &  23.1 &   0.1771 &   0.2340 &    127.5 & 3  A &   He I    & 3554.389   &  -2.4 & 3Po 1s.2p                           & 3D 1s.10d                           &  1.0 & **** &      2/0\\ 
 3560.616 &  20.3 &   0.0049 &   0.0065 &      7.3 & \nodata & \nodata & \nodata & \nodata & \nodata & \nodata & \nodata & \nodata & \nodata \\ 
 3562.932 &  20.7 &   0.0138 &   0.0182 &     11.5 & 4  A &   He I    & 3562.969   &   3.1 & 3Po 1s.2p                           & 3S 1s.10s                           &  2.0 &  1.0 &      1/0\\ 
 3574.775 &  13.5 &   0.0036 &   0.0047 &      7.2 & 6  B &   O  II   & 3574.845 : &   5.9 & 2D 2s2.2p2.(3P).3d                  & 2Po 2s2.2p2.(3P).5p                 &  1.5 &  1.5 &      2/0\\ 
 & & & & & 8  D &   N  II   & 3574.900 ? &  10.5 & 3P 2s.2p2.(4P).3d                   &
3Do 2s.2p2.(4P).5p                  &  2.0 &  2.0 &      2/0 \\
 3579.402 &  23.4 &   0.0103 &   0.0135 &      8.1 & \nodata & \nodata & \nodata & \nodata & \nodata & \nodata & \nodata & \nodata & \nodata \\
\enddata
\tablecomments{%
Columns: (1) Observed wavelength corrected for +68.6 \kms nebular
proper motion. (2) FWHM of line. (3) Observed intensity of line with
respect to the observed intensity of H$\beta$ (F(H$\beta$)=$1.31
\times 10^{-11}$ erg cm$^{-3}$ s$^{-1}$).  (4) De-reddened intensity
of line with respect to H$\beta$ (I(H$\beta$)=$2.87 \times 10^{-11}$
erg cm$^{-3}$ s$^{-1}$). (5) S/N of line; ``99999.0'' indicates
indeterminate. (6) EMILI IDI value/rank(A,B,C,D) where
applicable. ``*'' is IDI$<10$.  No IDI value/rank for manual IDs.  (7)
Source ion of line (8) Tabulated wavelength of transition.  ``:'' and
``*'' are as indicated in text. (9) Difference in wavelength between
observed and tabulated wavelength, $\delta V=
c*(\lambda-\lambda_o)/\lambda$ \kms. (10) Lower level
term/configuration.  Coupling schemes used are $LS$, $LK$, $J_1\,K$.
(11) Upper level term/configuration. (12) Lower level $J$ value,
``****'' indicates $J$ indeterminate (line from blended level) (13)
Upper level $J$ value (14) EMILI multiplet check statistics for $LS$
coupled levels only; $(A/B)$ where $A$ is number of lines expected and
$B$ the number considered found.  Not used for manual IDs.  
}

\end{deluxetable}

\clearpage

\begin{table}
\caption{Comparison of EMILI vs. Manual Line Identifications. \label{tbl4}}
\begin{center}
\begin{tabular}{crrrr}\tableline\tableline
 & \multicolumn{2}{c}{IC 418} & \multicolumn{2}{c}{Orion Nebula} \\
Rank &	\multicolumn{1}{c}{N} & \multicolumn{1}{c}{\%} & \multicolumn{1}{c}{N} & \multicolumn{1}{c}{\%} \\ \tableline  
A & 178	& 88 & 329 & 85 \\
B &   4	&  2 &  34 &  9 \\
C &   5	&  2 &  12 &  3 \\
D &   0 &  0 &   5 &  1 \\
  &  16 &  8 &   8 &  2 \\
Total &  203 & 	& 388 & \\ \tableline
\end{tabular}
\end{center}
\end{table}

\end{document}